\documentclass[conference]{IEEEtran}
\IEEEoverridecommandlockouts
\usepackage{cite}
\usepackage{amsmath,amssymb,amsfonts}
\usepackage{algorithmic}
\usepackage{graphicx}
\usepackage{textcomp}
\usepackage{xcolor}
\usepackage{url}
\usepackage{booktabs}
\usepackage{hyperref}

\def\BibTeX{{\rm B\kern-.05em{\sc i\kern-.025em b}\kern-.08em
    T\kern-.1667em\lower.7ex\hbox{E}\kern-.125emX}}
\begin{document}

\title{\huge\textbf{Underwater Source Detection and Classification for Signal-based Surveillance: Audio Dataset Curation and Cross-Domain Evaluation}}
% \vspace{-1.3ex}
\author{\IEEEauthorblockN{Quoc Thinh Vo and David Han}
\IEEEauthorblockA{\textit{Department of Electrical and Computer Engineering}, \textit{Drexel University}, Philadelphia, PA, USA \\
\{qv23, dkh42\}@drexel.edu}
}

% \author{\IEEEauthorblockN{1\textsuperscript{st} Given Name Surname}
% \IEEEauthorblockA{\textit{dept. name of organization (of Aff.)} \\
% \textit{name of organization (of Aff.)}\\
% City, Country \\
% email address or ORCID}

\maketitle

\begin{abstract}
Machine learning for underwater acoustics is constrained by the scarcity of publicly available labeled datasets. In contrast to air-acoustic domains, where large benchmarks enable rapid model development, underwater datasets are typically small and limited in acoustic diversity, restricting robust model training and cross-domain generalization. To help address this gap, we introduce a curated underwater audio dataset derived from an open-source maritime sound archive. The dataset contains over one thousand labeled audio segments across eight biologically and mechanically relevant acoustic classes, providing an additional resource for training models in data-limited underwater environments. Additionally, we establish a lightweight Convolutional Neural Network (CNN) baseline and propose a margin-enhanced loss with feature alignment to mitigate class confusion arising from data imbalance, acoustic similarity, and cross-domain mismatch. While the baseline achieves 96.35\% in-domain accuracy, evaluation on ShipsEar reveals substantial domain shift; the proposed feature alignment improve zero-shot ship detection by 42.60\%, demonstrating stronger robustness under distribution mismatch. We further release a transparent curation pipeline and reproducible benchmark to support future research on imbalance mitigation, domain adaptation, and data-efficient underwater acoustic classification.
\end{abstract}


\section{Introduction}
\label{sec:intro}

\begin{figure*}
    \centering
\includegraphics[width=0.91\linewidth]{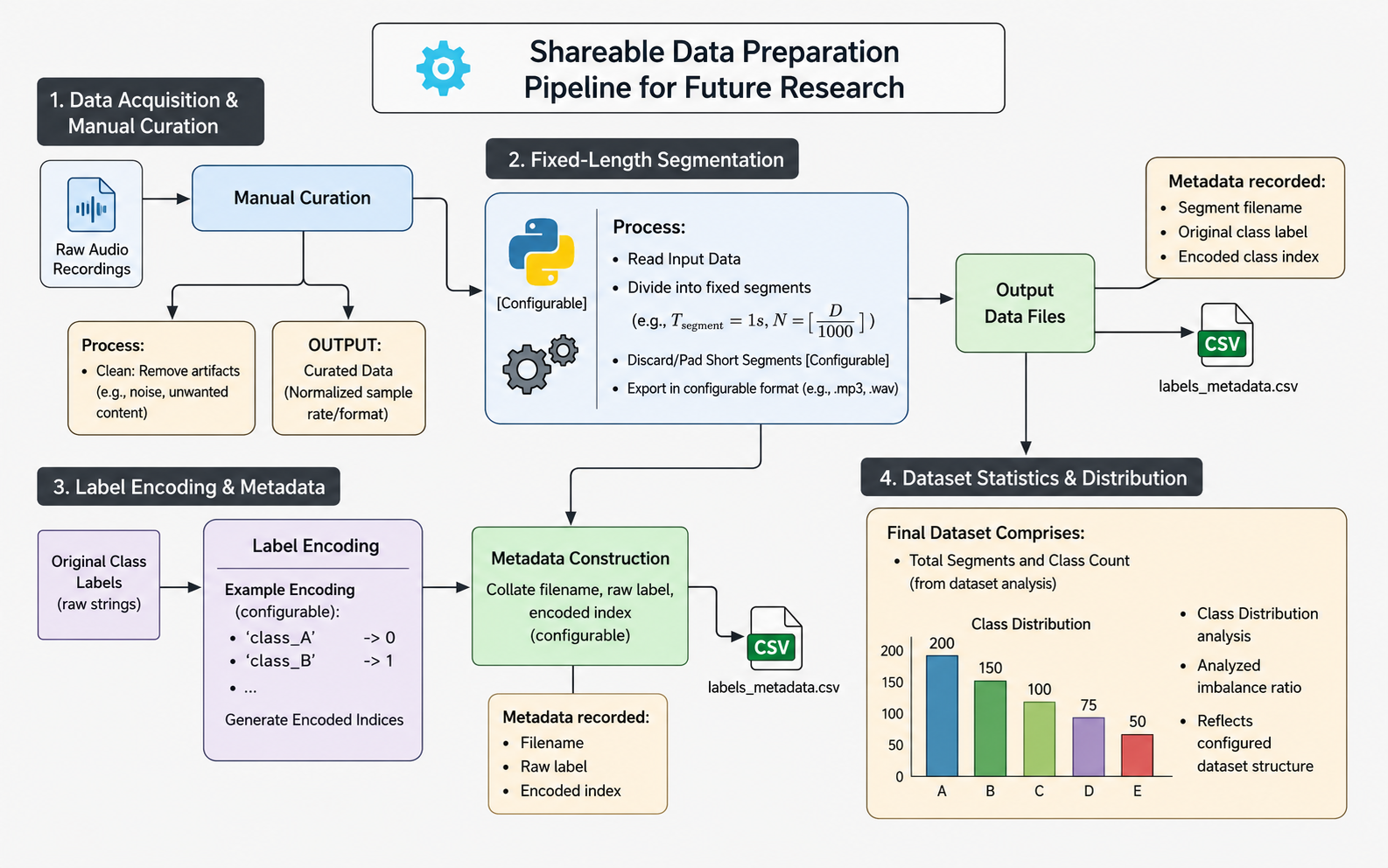}
    \caption{Data Preparation Pipeline.}
    \label{data_pipeline}
\end{figure*}

Machine learning (ML) has become a key technique for underwater acoustic analysis, supporting applications such as sensor monitoring, vessel detection, maritime surveillance, and sound source classification and localization \cite{Muthu2025Underwater, Feng2022Transformer, VTransformer, Zhang2022ModulationRNN,
vo2025adaptive, chowdhury2025multi, Guo2022UnderwaterCNN,  VoICASSP26}. Despite these advances, progress remains limited by the scarcity of publicly available underwater datasets. In contrast, air-acoustic research benefits from widely used benchmarks such as ESC50 (2,000 clips) \cite{piczak2015esc}, GTZAN (1,000 clips) \cite{tzanetakis2002musical}, UCF101 (2,840 clips) \cite{soomro2012ucf101}, and ActivityNet (1,760 clips) \cite{caba2015activitynet}, which have enabled rapid development of deep learning models. Underwater datasets, however, remain comparatively small and fragmented \cite{parsons2022sounding}. Data collection in marine environments is difficult due to harsh conditions, specialized instrumentation, and the logistical complexity of deploying and maintaining recording systems \cite{Chen2021Achieving}. Consequently, existing datasets often contain limited acoustic diversity and uneven class distributions. Models trained on such restricted domains frequently fail to generalize across environments \cite{Kataria2025Domain-Robust}, as variations in recording location, sensor characteristics, environmental conditions, and acoustic propagation introduce substantial domain shift. Furthermore, underwater recordings commonly include overlapping biological and anthropogenic sounds, strong background noise, and class imbalance \cite{Xie2024Unraveling}. The absence of standardized preprocessing \cite{pessanha2023expansion}, segmentation, and labeling pipelines also limits reproducibility and fair benchmarking \cite{ren2019feature}. These challenges highlight the need for curated datasets and transparent data-processing pipelines that support robust evaluation and cross-domain learning in underwater acoustic machine learning.

To address these limitations, we introduce a curated underwater audio dataset (USS8) derived from an open-source maritime sound archive \cite{maritime}, intended for academic and non-commercial use. The recordings were manually curated to remove unrelated content and segmented into non-overlapping 1-second clips across eight acoustic categories with structured metadata. We establish a reproducible benchmark using a lightweight Tiny-CNN \cite{wei2018quantization} trained on log mel-spectrogram representations. In addition, we investigate class imbalance and inter-class acoustic similarity through a margin-enhanced loss formulation that improves separation between confusing categories. To assess robustness, we further evaluate cross-domain transfer on the ShipsEar \cite{santos2016shipsear} dataset, examining the impact of domain shift and the effectiveness of margin-based decision-boundary regularization. The contributions of our work are summarized as follows:
% \vspace{-\parskip}
% \begingroup\setlength{\abovedisplayskip}{5pt}\setlength{\belowdisplayskip}{3pt}
\begin{itemize}
    \item \textbf{Baseline and Imbalance Study:} A lightweight CNN benchmark with margin-based loss analysis.
    \item \textbf{Cross-Domain Evaluation:} Empirical assessment of transfer performance on the ShipsEar dataset.
    \item \textbf{Data Preparation Pipeline:} A modular audio curation and segmentation pipeline with open-source code for preparing datasets from public audio archives \footnote{\url{https://github.com/qtvo93/data-pipeline-avss}}. The pipeline enables reproduction of the curated underwater 8-class dataset with standardized segmentation and labeling; however, the curated dataset itself cannot be redistributed due to licensing restrictions.
\end{itemize}
% \endgroup

\section{Methodology}
\label{sec:method}

\subsection{Audio Curation and Segmentation Procedure}

\subsubsection{Data Acquisition and Manual Curation}

The raw audio recordings were obtained from~\cite{maritime}. 
Because the original recordings contained introductory material, silence periods, and unrelated acoustic content, each file was manually inspected prior to segmentation. Audio portions not corresponding to the labeled target events were removed to ensure strict alignment between waveform content and ground-truth annotations. This manual curation step reduces label noise and prevents contamination of the training data with irrelevant signals. Additionally, all audio files were inspected to ensure an original uniform format. This guarantees uniform spectral resolution during subsequent feature extraction and model training.

\subsubsection{Fixed-Length Segmentation}

To standardize input duration for model training, each curated audio track was segmented into non-overlapping 1-second intervals using the \texttt{pydub} library \cite{pydub}. Let $x(t)$ denote the continuous-time audio waveform. The digital recording was divided into contiguous segments of fixed duration:

\begin{equation}
T_{\text{segment}} = 1~\text{second}.
\end{equation}

For an audio file of total duration $D$ milliseconds, segmentation yields

\begin{equation}
N = \left\lfloor \frac{D}{1000} \right\rfloor
\end{equation}

segments $\{x_0, x_1, \dots, x_{N-1}\}$. 

Any residual segment shorter than one second was discarded to maintain consistent input dimensionality across all samples. Each extracted segment was exported as an independent \texttt{.mp3} file to facilitate modular data loading and reproducibility.

\subsubsection{Label Encoding and Metadata Construction}

Each audio track was associated with a predefined semantic label (e.g., \textit{torpedo}, \textit{whale}, \textit{ships}). A categorical mapping dictionary was defined to convert string labels into integer class indices for supervised training: $\text{class\_dict} = \{\text{whale}:0,\ \text{fish}:1,\ \text{unidentified}:2,\ \text{physical}:3,\ \text{ships}:4,\ \text{communications}:5,\ \text{sonar}:6,\ \text{torpedo}:7\}$.

For each segment, the following metadata were recorded:

% \begin{equation}
% \text{class\_dict} =
% \begin{cases}
% \text{whale} \rightarrow 0 \\
% \text{fish} \rightarrow 1 \\
% \text{unidentified} \rightarrow 2 \\
% \text{physical} \rightarrow 3 \\
% \text{ships} \rightarrow 4 \\
% \text{communications} \rightarrow 5 \\
% \text{sonar} \rightarrow 6 \\
% \text{torpedo} \rightarrow 7
% \end{cases}
% \end{equation}

% \vspace{-\parskip}
% \begingroup\setlength{\abovedisplayskip}{4pt}\setlength{\belowdisplayskip}{2pt}
\begin{itemize}
    \item Segment filename
    \item Original class label (string)
    \item Encoded class index (integer)
\end{itemize}
% \endgroup

All metadata entries were stored in a structured CSV file (\texttt{labels\_metadata.csv}) to support reproducibility and downstream data loading.

\subsubsection{Dataset Statistics and Class Distribution}

After manual curation and fixed-length segmentation, the final dataset comprised 1{,}099 non-overlapping 1-second audio segments across eight predefined acoustic classes. This segmentation strategy ensures consistent input dimensionality, reduced label ambiguity, and compatibility with batch-based ML pipelines for supervised training and evaluation. The class distribution is moderately imbalanced, with the largest class containing 273 samples (24.84\%) and the smallest class containing 52 samples (4.73\%), resulting in an imbalance ratio (max/min) of 5.25. 
Such imbalance reflects the natural occurrence variability of underwater acoustic events and is considered during model training and evaluation. The detailed class distribution is summarized in Table~\ref{tab:dataset_distribution}.

\begin{table}[t]
\small
\caption{Class distribution of the curated underwater dataset.}
\label{tab:dataset_distribution}
\centering
\setlength{\tabcolsep}{4pt}
\renewcommand{\arraystretch}{1.05}
\begin{tabular}{llll}
% \begin{tabular}{@{}l c S[table-format=1.4] S[table-format=2.2]@{}}
\toprule
\textbf{Class ID} & \textbf{Label} & \textbf{\# Segments} & \textbf{Percentage (\%)} \\
\hline
0 & Whale & 273 & 24.84 \\
1 & Fish & 99 & 9.01 \\
2 & Unidentified & 94 & 8.55 \\
3 & Physical & 122 & 11.10 \\
4 & Ships & 81 & 7.37 \\
5 & Communications & 200 & 18.20 \\
6 & Sonar & 178 & 16.20 \\
7 & Torpedo & 52 & 4.73 \\
\bottomrule
\end{tabular}
\end{table}

\subsection{Baseline Benchmark}

To demonstrate the practical utility of the dataset and establish a reproducible benchmark, we train a lightweight Tiny-CNN for 8-class acoustic classification. The architecture is deliberately kept simple to minimize overfitting on the limited dataset and to ensure that the benchmark can be easily reproduced and extended by future work. The architecture is summarized in Table~\ref{tab:Tiny-CNN_arch}.

\subsubsection{Feature Extraction}

\begin{figure}
    \centering
\includegraphics[width=1\linewidth]{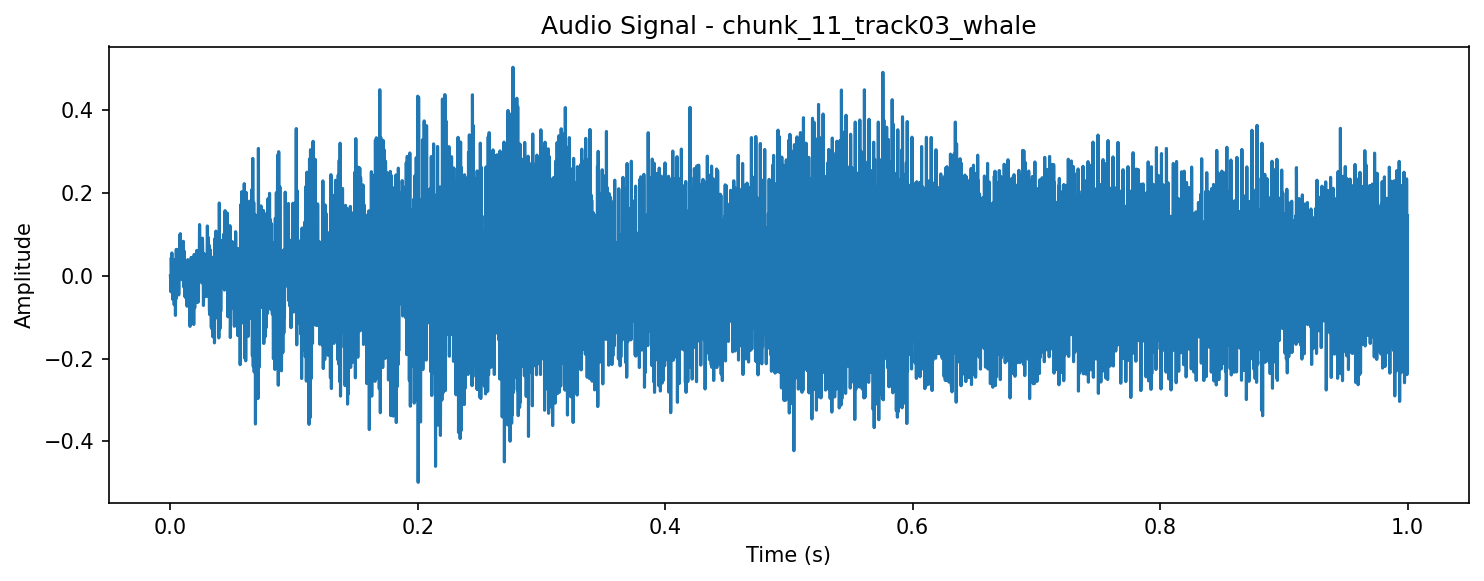}
\caption{Example waveform of a 1-second audio segment.}
    \label{sample_audio}
\end{figure}

\begin{figure}
    \centering
\includegraphics[width=1\linewidth]{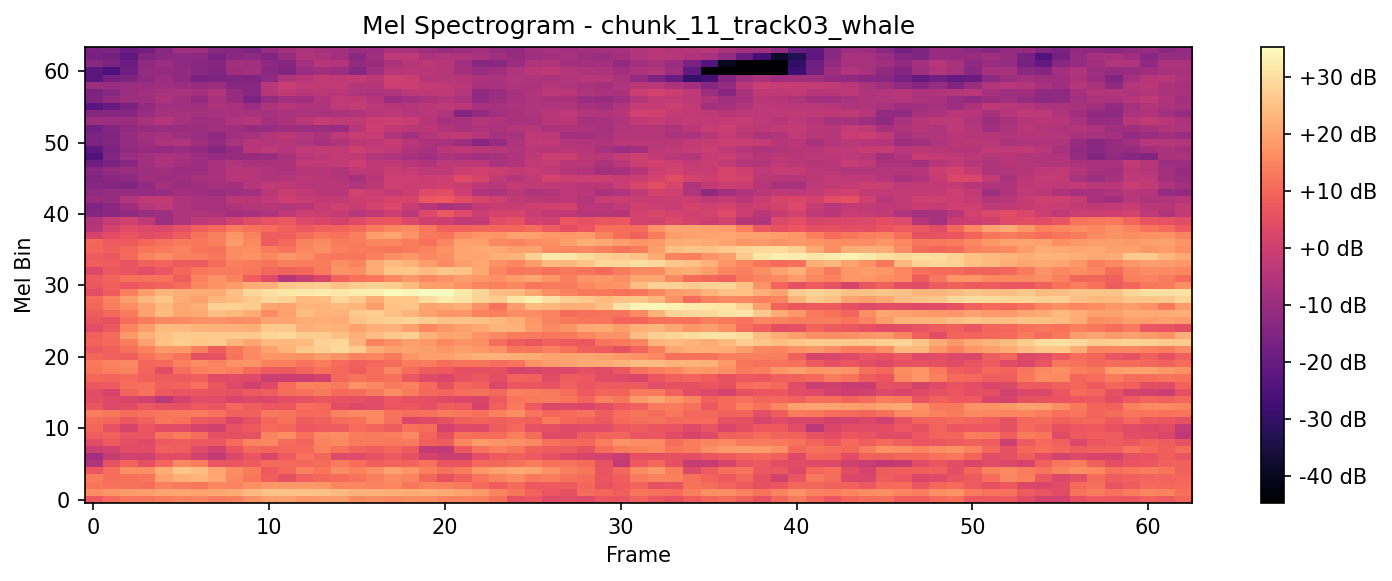}
\caption{Mel-spectrogram representation of the corresponding 1-second audio segment.}
    \label{sample_spectrogram}
\end{figure}

Each 1-second waveform segment is converted to a log Mel-spectrogram using a Mel-scaled STFT with $n_{\text{fft}}=1024$, hop length $=256$, sampling rate $=16{,}000$~Hz, and $n_{\text{mels}}=64$, following \cite{krause2024sound, Vo_DU_task3a_report}.

For a waveform $x(t)$ (Figure \ref{sample_audio}), the Mel-spectrogram (Figure \ref{sample_spectrogram}) is computed as:

\begin{equation}
\mathbf{M} = \text{MelSpectrogram}(x; n_{\text{fft}}=1024, \text{hop}=256),
\end{equation}

followed by logarithmic amplitude scaling:

\begin{equation}
\mathbf{M}_{\text{dB}} = 10 \log_{10}(\mathbf{M}),
\end{equation}

implemented using an amplitude-to-decibel transformation with the dynamic range clipped to $[-80, 80]$~dB. To improve training stability and reduce sample-level energy variation, each log Mel-spectrogram is standardized as:

\begin{equation}
\mathbf{\hat{M}} = \frac{\mathbf{M}_{\text{dB}} - \mu}{\sigma + \epsilon},
\end{equation}

where $\mu$ and $\sigma$ are the mean and standard deviation of the sample, and $\epsilon = 10^{-6}$ ensures numerical stability. The normalized Mel-spectrograms are then used as input to the Tiny-CNN.

\subsubsection{Training Setup}

The dataset is randomly split into 716 training, 164 validation, and 219 test samples, for a total of 1,099 segments. This split is fixed across all experiments to ensure fair comparison and reproducibility. The model is trained with cross-entropy loss and early stopping based on validation performance. Deterministic training is used for reproducibility, and all experiments are conducted on an NVIDIA GeForce RTX 3090 GPU and an AMD Ryzen Threadripper PRO 3955WX CPU. Detailed classification results are reported in Table~\ref{tab:baseline_metrics}, where the Support column indicates the number of test samples.

\begin{table}[t]
\small
\caption{Tiny-CNN architecture.}
\label{tab:Tiny-CNN_arch}
\centering
\setlength{\tabcolsep}{4pt}
\renewcommand{\arraystretch}{1.05}
\begin{tabular}{lll}
\toprule
\textbf{Layer} & \textbf{Configuration} & \textbf{Output Channels} \\
\hline
Conv2D + BN + ReLU & $3 \times 3$ kernel & 32 \\
MaxPool2D & $2 \times 2$ & 32 \\
Conv2D + BN + ReLU & $3 \times 3$ kernel & 64 \\
MaxPool2D & $2 \times 2$ & 64 \\
Conv2D + BN + ReLU & $3 \times 3$ kernel & 128 \\
Global Avg Pool & $1 \times 1$ & 128 \\
Fully Connected & Linear & 8 classes \\
\bottomrule
\end{tabular}
\end{table}

\begin{table}[t]
\small
\caption{Baseline Tiny-CNN performance on the test set.}
\label{tab:baseline_metrics}
\centering
\setlength{\tabcolsep}{4pt}
\renewcommand{\arraystretch}{1.05}
\begin{tabular}{lllll}
\toprule
\textbf{Class} & \textbf{Support} & \textbf{Precision} & \textbf{Recall} & \textbf{F1} \\
\hline
Whale & 61 & 0.967 & 0.951 & 0.959 \\
Fish & 13 & 1.000 & 1.000 & 1.000 \\
Unidentified & 16 & 0.941 & 1.000 & 0.970 \\
Physical & 21 & 0.909 & 0.952 & 0.930 \\
Ships & 14 & 1.000 & 0.857 & 0.923 \\
Communications & 40 & 0.927 & 0.950 & 0.938 \\
Sonar & 42 & 1.000 & 1.000 & 1.000 \\
Torpedo & 12 & 1.000 & 1.000 & 1.000 \\
\midrule
Macro Avg & 219 & 0.968 & 0.964 & 0.965 \\
Weighted Avg & 219 & 0.964 & 0.963 & 0.963 \\
Overall Accuracy & 219 & \multicolumn{3}{c}{0.9635} \\
\bottomrule
\end{tabular}
\end{table}

\subsection{Cross-Domain Evaluation on ShipsEar}

To evaluate cross-domain generalization, the pretrained 8-class model is tested on ShipsEar~\cite{santos2016shipsear}, a dataset of vessel and ambient noise recordings collected near the port of Vigo on the Spanish Atlantic coast. ShipsEar contains 90 recordings spanning 12 classes, including 11 vessel categories and one natural ambient-noise class. For consistency, all recordings are segmented into non-overlapping 1-second clips. The original labels are consolidated into a binary task: \textit{ships} versus \textit{ambient noise}. Predictions from the pretrained 8-class model are mapped to this 2-class setting by treating class ID 4 (Ships) as the positive ship class.

The evaluation set is highly class-imbalanced, comprising 11,300 segments, including 10,160 ship segments and only 1,140 ambient-noise segments. This imbalance poses a significant challenge, particularly for models not designed to handle skewed class distributions. As shown in Table~\ref{tab:shipsear_results}, the baseline model trained with standard cross-entropy exhibits poor cross-domain performance, achieving only 5.91\% ship detection rate and 15.31\% overall accuracy. This degradation is primarily attributed to the combined effects of severe class imbalance and domain shift between the source training data and the ShipsEar dataset. Consequently, the model tends to favor more familiar source-domain classes, leading to frequent misclassification of ship segments despite their majority presence in the evaluation set.

\begin{table*}[t]
\small
\caption{Cross-domain ship detection performance on ShipsEar - zero-shot and test-time feature alignment applied.}
\label{tab:shipsear_results}
\centering
\setlength{\tabcolsep}{4pt}
\renewcommand{\arraystretch}{1.1}
\begin{tabular}{lccccc}
\toprule
\textbf{Model Tiny-CNN} 
& \textbf{Ship Det. Rate (\%)} 
& \textbf{Ship F1} 
& \textbf{Overall Acc. (\%)} 
& \textbf{Balanced Acc. (\%)} 
& \textbf{\# Segments} \\
\midrule
Standard CE zero-shot
& 5.91 
& 0.111 
& 15.31 
& 52.51 
& 11{,}300 \\

Class-weighted CE
& 16.22 
& 0.277
& 23.68
& 53.19
& 11{,}300 \\

CE-PlusPairMargin
& 29.43 
& 0.448 
& 34.74 
& 55.77 
& 11{,}300 \\

CE-PlusPairMargin + Feature Alignment
& \textbf{48.51} 
& \textbf{0.644} 
& \textbf{51.86} 
& \textbf{65.09} 
& 11{,}300 \\
\bottomrule
\end{tabular}
\end{table*}

\subsection{Margin-Enhanced Loss and Feature Alignment for Class Confusion Mitigation}

To improve robustness to both class imbalance and domain shift, we introduce two complementary components: a margin-enhanced loss during training and a lightweight feature-statistics alignment step during inference.

\begin{figure*}
    \centering
    \includegraphics[width=0.9\linewidth]{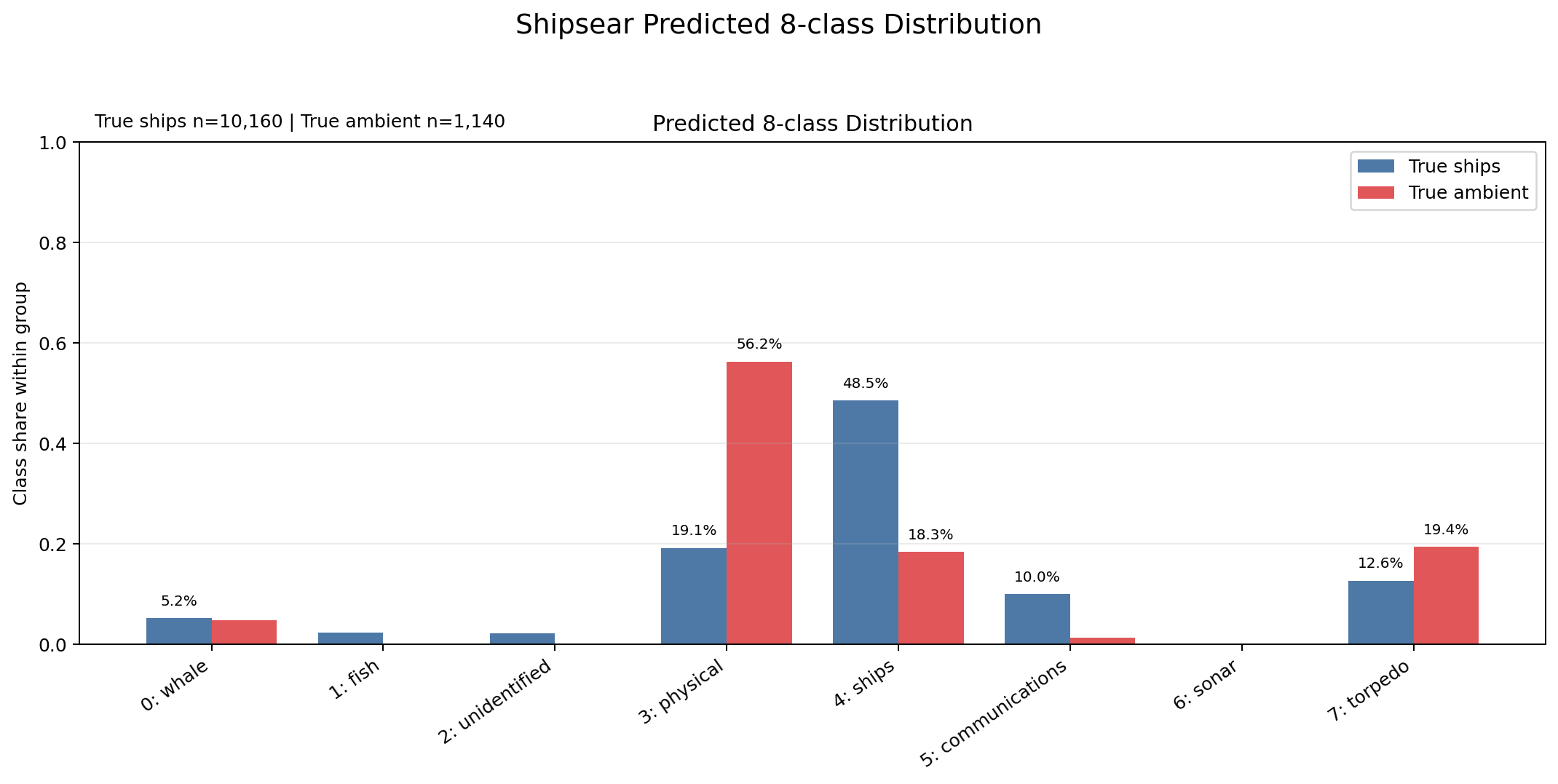}
    \caption{ShipsEar classification report illustrating how ship and ambient noise being classified.}
\label{shipsear_report_classification_flow}
\end{figure*}

We incorporate class-dependent weighting into the cross-entropy loss \cite{fernando2021dynamically} to reduce class imbalance. Let $n_i$ denote the number of samples in class $i$. Following an inverse-frequency strategy, the class weights are defined as

\begin{equation}
w_i = \frac{\frac{1}{n_i}}{\frac{1}{C}\sum_{j=1}^{C}\frac{1}{n_j}}
\end{equation}

where $C$ is the number of classes. This normalization preserves an average weight of one while increasing the contribution of minority classes, thereby reducing bias toward dominant classes such as whale and sonar. Results show that ship samples (class 4) are frequently confused with whale (class 0) and sonar (class 6), primarily due to class imbalance and spectral similarity.

We introduce \textit{CE-PlusPairMargin}, which augments the weighted cross-entropy with margin-based logit constraints inspired by the Adaptive-Weighted-Loss of Roy et al.~\cite{Roy2024Margin-Aware}. For ship samples, we enforce a minimum margin between the ship logit and the competing whale and sonar logits:

\begin{equation}
z_4 - z_0 \geq m, \quad z_4 - z_6 \geq m
\end{equation}

where $z_i$ denotes the logit for class $i$ and $m$ is the margin parameter. The final loss is

\begin{equation}
\mathcal{L} = \mathcal{L}_{CE}(\mathbf{w}) + \lambda \mathcal{L}_{margin}
\end{equation}

where $\mathcal{L}_{CE}(\mathbf{w})$ is the class-weighted cross-entropy and $\lambda$ controls the margin regularization strength.

Because cross-domain degradation also arises from feature mismatch, we further apply a feature-statistics alignment step at inference time. Let $\mathbf{X}$ denote the target-domain log-mel spectrogram. Using the source-domain statistics $(\mu_s, \sigma_s)$ and target-domain statistics $(\mu_t, \sigma_t)$, the aligned representation $\tilde{\mathbf{X}}$ is computed as

\begin{equation}
\tilde{\mathbf{X}} =
\left(
\frac{\mathbf{X} - \mu_t}{\sigma_t + \epsilon}
\right)\sigma_s + \mu_s
\end{equation}

where $\epsilon$ is a small constant for numerical stability. The aligned and original features are blended using an interpolation parameter $\alpha$:

\begin{equation}
\mathbf{X}_{align} =
(1-\alpha)\mathbf{X} + \alpha \tilde{\mathbf{X}}
\end{equation}

To avoid excessive shifts, the aligned features are clipped around the source statistics:

\begin{equation}
\mathbf{X}_{align} =
\text{clip}\left(
\mathbf{X}_{align},
\mu_s - k\sigma_s,
\mu_s + k\sigma_s
\right)
\end{equation}

Finally, per-sample normalization is applied:

\begin{equation}
\mathbf{X}_{norm} =
\frac{\mathbf{X}_{align} - \mu(\mathbf{X}_{align})}
{\sigma(\mathbf{X}_{align}) + \epsilon}
\end{equation}

This alignment acts as a lightweight domain adaptation mechanism and requires no retraining. While the margin-enhanced loss yields limited gain on the near-saturated 8-class in-domain task, it substantially improves cross-domain ship detection. As shown in Table~\ref{tab:shipsear_results}, ship detection increases from $5.91\%$ under standard CE to $29.43\%$ with margin-enhanced training, and further to $48.51\%$ after feature alignment as shown in Table \ref{tab:shipsear_results} and Figure \ref{shipsear_report_classification_flow}.

\section{Results and Analysis}

\subsection{Evaluation Metrics}

Let $TP$, $TN$, $FP$, and $FN$ denote the number of true positives, true negatives, false positives, and false negatives, respectively.

Ship detection rate is defined as the recall of the ship class:

\begin{equation}
\text{Recall}_{\text{ship}} = \frac{TP}{TP + FN}.
\end{equation}

The F1 score combines precision and recall:

\begin{equation}
F_1 = \frac{2PR}{P + R},
\end{equation}

where

\begin{equation}
P = \frac{TP}{TP + FP}, \quad R = \frac{TP}{TP + FN}.
\end{equation}

Balanced accuracy accounts for class imbalance by averaging recall across classes:

\begin{equation}
\text{Balanced Accuracy} =
\frac{\text{Recall}_{\text{ship}} + \text{Recall}_{\text{ambient}}}{2}.
\end{equation}

Overall accuracy is defined as

\begin{equation}
\text{Accuracy} = \frac{TP + TN}{TP + TN + FP + FN}.
\end{equation}

\subsection{In-Domain Performance Analysis}

Table~\ref{tab:baseline_metrics} summarizes in-domain results on the USS8 test set. The Tiny-CNN achieves 96.35\% overall accuracy with macro and weighted F1-scores of 0.965 and 0.963, indicating stable performance despite dataset imbalance. Perfect precision and recall are observed for \textit{Fish}, \textit{Sonar}, and \textit{Torpedo}, suggesting distinct spectral characteristics. In contrast, the \textit{Ships} class achieves perfect precision (1.000) but lower recall (0.857), indicating that some ship samples are confused with acoustically similar categories. Slightly lower F1-scores for \textit{Physical} and \textit{Communications} also reflect residual inter-class overlap. Overall, the results confirm that the curated dataset is learnable with a lightweight architecture while revealing early signs of ship-class confusion.

\subsection{Domain Shift Analysis}

Despite strong in-domain performance, cross-domain evaluation under standard CE training shows severe degradation, indicating substantial distribution shift between USS8 and ShipsEar. Differences in recording hardware, environmental conditions, vessel operating modes, and background noise profiles introduce spectral and temporal mismatch. In addition, USS8 contains fewer vessel types than ShipsEar, which constrains the model's ability to learn sufficiently diverse and robust vessel-specific features, thereby further increasing domain discrepancy. To analyze the impact of feature alignment, we compare predictions before and after alignment. Among the 11{,}300 evaluated segments, 8{,}802 predictions (77.89\%) change class labels, indicating that the alignment significantly reshapes the feature distribution perceived by the classifier. The most frequent prediction transitions are summarized as:

\begin{equation}
\text{TopTransitions} =
\begin{cases}
0 \rightarrow 4 : 2347 \\
0 \rightarrow 3 : 1542 \\
0 \rightarrow 7 : 746 \\
4 \rightarrow 3 : 520 \\
0 \rightarrow 5 : 486 \\
3 \rightarrow 5 : 373 \\
2 \rightarrow 4 : 351 \\
4 \rightarrow 7 : 301 \\
6 \rightarrow 4 : 295 \\
3 \rightarrow 7 : 267
\end{cases}
\end{equation}

Inspection of class transitions shows that the most frequent correction is whale-to-ship ($0 \rightarrow 4$), affecting 2{,}347 segments. This directly addresses the dominant confusion pattern from the results analysis. Additional transitions such as $2 \rightarrow 4$ (351 segments) and $6 \rightarrow 4$ (295 segments) further contribute to improved ship detection. Other frequent changes include $0 \rightarrow 3$ (1{,}542 segments) and $0 \rightarrow 7$ (746 segments), indicating that the alignment globally adjusts the feature distribution rather than targeting a single class boundary.

\subsection{Discussion}

Despite achieving 96.35\% in-domain accuracy, the CE-trained model detects only 5.91\% of ships in the cross-domain setting, showing that strong closed-set performance does not necessarily translate to generalization. The margin-enhanced loss and feature alignment improve cross-domain detection to 48.51 \%, indicating that structured inter-class separation improves robustness to acoustically overlapping categories. However, the remaining gap suggests these methods alone cannot fully address domain mismatch. Moreover, although computationally efficient, the Tiny-CNN may lack sufficient capacity to model the complex temporal dynamics present in real-world maritime recordings.

\section{Conclusion}

While margin-enhanced training improves classification performance, several limitations remain. Variability in vessel characteristics, operational behavior, and environmental conditions is not fully captured by the current dataset, potentially limiting cross-domain generalization. Additionally, the transfer evaluation is based on direct model reuse, without the use of explicit domain adaptation or more sophisticated feature-alignment methods. As reflected in the cross-domain performance gap, distribution mismatch between datasets remains a major challenge. Future work will focus on expanding ship diversity in the dataset, incorporating domain adaptation strategies such as feature alignment or limited target-domain fine-tuning, and exploring self-supervised or contrastive pretraining to learn more transferable acoustic representations. These efforts aim to improve robustness and support more reliable deployment in real-world underwater surveillance scenarios.

\section*{Acknowledgment}
The work on this paper was supported by the Office of Naval Research (Grant No. N00014-21-1-2790).

% The work on this paper was supported by the Office of Naval Research. Grant No. N00014-21-1-2790.

% \fontsize{8.26pt}{8.3pt}\selectfont

% References should be produced using the bibtex program from suitable
% BiBTeX files (here: strings, refs, manuals). The IEEEbib.bst bibliography
% style file from IEEE produces unsorted bibliography list.
% -------------------------------------------------------------------------
% \bibliographystyle{IEEEbib}
% \bibliography{strings,refs}
\bibliographystyle{IEEEtran}
\bibliography{refs}

\end{document}